\documentclass[prd,onecolumn,aps,floats,superscriptaddress,floatfix,
showpacs,showkeys,amsmath,amssymb,nofootinbib]{revtex4}

\usepackage{dcolumn}% Align table columns on decimal point
\usepackage{bm}% bold math
\usepackage{txfonts}% bold math
\usepackage{pxfonts}% bold math
\usepackage{graphicx}
\usepackage{epsfig} 
\usepackage{psfrag}
\usepackage{times}

\begin{document}
\title{Implications of nonlinearity for spherically symmetric
accretion} 
\author{Sourav Sen}\email{souravsen@iiserkol.ac.in}
\affiliation{Department of Physical Sciences,
Indian Institute of Science Education and Research
(IISER Kolkata), Mohanpur Campus, PO: BCKV Campus Main Office,
Mohanpur 741252, West Bengal, India}
\author{Arnab K. Ray}\email{arnab.kumar@juet.ac.in}
\affiliation{Department of Physics, Jaypee University
of Engineering \& Technology, %A-B Road,
Raghogarh, Guna 473226, Madhya Pradesh, India}
\date{\today}

\begin{abstract}
We subject the steady solutions of a spherically symmetric 
accretion flow to a time-dependent radial perturbation. The
equation of the perturbation 
includes nonlinearity up to any arbitrary order, and  
bears a form that is very similar to the 
metric equation of an analogue acoustic black hole. Casting
the perturbation as a standing wave on subsonic solutions,
and maintaining nonlinearity in it up to the second order, 
we get the time-dependence 
of the perturbation in the form of a Li\'enard system. A 
dynamical systems analysis of the Li\'enard system reveals 
a saddle point in real time, with the implication that 
instabilities
will develop in the accreting system when the perturbation 
is extended into the nonlinear regime. The instability of 
initial subsonic states also adversely affects the temporal 
evolution of the flow towards a final and stable transonic state. 
\end{abstract}


\maketitle

\section{Introduction}
\label{sec1}
The classical model of 
steady spherically symmetric accretion~\cite{bon52}
is a mathematical problem of conservative and 
compressible hydrodynamics. This model has acquired a 
paradigmatic status in studies on astrophysical accretion, 
which, as a fluid flow, falls under the general class of 
nonlinear dynamics. The mathematical description of such fluid
systems involves a momentum balance equation (with gravity 
as an external force in accretion), the continuity equation 
and a polytropic equation of state~\cite{fkr02}. Set in full 
detail, the condition of momentum conservation in a fluid is 
a balance of dynamic effects, nonlinear effects and the effects 
of the pressure in a continuum system~\cite{ll87}. 
Some early studies on astrophysical accretion 
considered only the interplay between dynamics and 
nonlinearity (see~\cite{bon52} and references therein), but 
in present studies, it is customary to ignore the dynamics 
and instead consider the effects of pressure, in what becomes
a stationary flow. In either case, however, nonlinearity endures.

From the plethora of mathematical solutions of the stationary
spherically symmetric compressible fluid flow, the ones of 
physical relevance
are identified to be locally subsonic very far away from the 
accretor. Within the class of inflows obeying this outer boundary
condition, there is an infinitude of globally subsonic solutions,
along which a fluid element may reach the accretor with a low 
subsonic velocity. For the same outer boundary condition,  
a single critical solution stands out in a class by itself, 
and allows matter to reach the accretor with a high supersonic 
velocity, crossing the sonic horizon along the way. This 
is the unique transonic~\citet{bon52} accretion solution.
The exact fashion in which accreting matter reaches the 
accretor is related to the inner boundary condition of the 
inflow problem. If the accretor is a black hole,
the infall process must be transonic~\cite{nt73,shateu83}.
This is because a black hole has an event horizon instead of 
a physical surface, and thus precludes all possibility of a 
pressure build-up at small radii, which could otherwise have 
dominated over the free-fall conditions close to the accretor.  
The situation is not so clearly understood if the 
accretor has a hard surface like a neutron star or a white dwarf. 
For such an accretor, it is supposed that the accumulated matter 
will build up pressure near the surface and cause the supersonic 
flow to be shocked down to subsonic levels, although for a 
neutron star, all accreted matter might be 
``vacuum cleaned" away efficiently, making it easier 
for the flow to remain supersonic~\cite{pso80}. Evidently
then, the question of the inner 
boundary condition and an inflow trajectory 
in relation to it, is by no means a trivial one. 
Nevertheless, working with the stationary problem
itself,~\citet{bon52} invoked the physical criteria
of the maximization of the accretion rate and the minimization 
of the total energy of the flow, to propose that the transonic 
solution would be the one selected by a fluid element to 
reach the accretor from a distant outer boundary. 
\citet{bon52}, however, left a definitive conclusion
regarding the realizability of the transonic solution  
to its stability. 

The trouble with the transonic solution in the stationary 
regime is that its realizability is extremely vulnerable 
to even an infinitesimal deviation from the precisely needed
boundary condition to generate the solution~\citep{rb02}. 
This difficulty may be overcome by considering the  
temporal evolution of global solutions towards the transonic 
state~\citep{rb02,rrbon07}, but there is no analytical 
formulation to solve the nonlinear partial differential 
equations governing the temporal evolution of the flow. 
So, much of all time-dependent studies in spherically 
symmetric accretion is perturbative and linearized
in character, although some non-perturbative
studies are also known~\citep{rb02,rrbon07}. 
The commonly accepted view to have emerged from the linearized
approach is that perturbations on the flow do not
produce any linear mode with an amplitude that gets amplified
in time~\citep{gai06}, and that the perturbative method does
not indicate the primacy of any particular class of 
solutions~\citep{gar79}. This is as far as could be said,  
working in the linear regime, but our general experience 
of any nonlinear system is that an understanding gained 
about it under linearized conditions
can scarcely be imposed on circumstances dominated
by nonlinearity. In this work we attempt to bridge the gap. 

First we adopt a time-dependent radial 
perturbation scheme implemented originally 
by~\citet{pso80} and retain all orders of nonlinearity
in the resulting equation of the perturbation.
A most striking feature of this equation 
is that even on accommodating nonlinearity in 
full order, it conforms to the structure of the metric equation 
of a scalar field in Lorentzian geometry (Section~\ref{sec3}). 
This fluid analogue
(an ``acoustic black hole"), emulating many features of a 
general relativistic black hole, is a matter of continuing 
interest in fluid mechanics from diverse points of 
view~\citep{monc80,un81,jacob91,un95,vis98,bil99,su02,das04,sbr05,
us05,vol05,vol06,rbcqg07,rbpla07,rrbon07,ncbr07,dbd07,macmal08,
blv11,rob12,sbbr13}.   
Then we apply our nonlinear equation of the perturbation to 
study the stability of globally subsonic stationary solutions 
under large-amplitude time-dependent perturbations. Our 
motivation to do so lies in certain dynamic features of the 
flow. For the non-perturbative 
time evolution of the accreting system, 
the initial condition of the evolution
is a globally subsonic state, with gravity subsequently 
driving the system to a transonic state, sweeping through an
infinitude of intermediate subsonic states. So, to ensure an 
unhindered temporal convergence to a stable transonic 
trajectory, the stability of the subsonic states 
is crucial. In a numerical study, an instability 
in the subsonic states was observed by~\citet{sb78}, who 
were consequently of the view that the subsonic flows would
quickly change into the transonic flow. 
Our nonlinear perturbative analysis does agree 
with the fact that there is an instability in the subsonic 
states, but short of an exact non-perturbative analysis of 
all the nonlinear flow equations concerned (for which 
there exists no analytical prescription yet), 
it would be hasty to claim that the transonic state is indeed the
final stable attractor state for the unstable subsonic states.
The feasible analytical alternative, therefore, is to study 
the behaviour of the system under progressively higher orders 
(nonlinear orders) of time-dependence in the perturbative 
approach. We truncate all orders of nonlinearity
beyond the second order in our equation of 
the perturbation. Following this, we integrate out the spatial 
dependence of the perturbation with the help of well-defined 
boundary conditions on globally subsonic 
flows~\citep{pso80,td92}. 
After this, we extract the time-dependent part of the perturbation 
and note with intrigue that it has the mathematical form of a 
Li\'enard system~\citep{stro,js99} (Section~\ref{sec4}). 
On applying the standard analytical
tools of dynamical systems to study the equilibrium features 
of this Li\'enard system, we discover the existence of a saddle 
point in real time, whose implication is that the stationary 
background solutions will be unstable, if the 
perturbation is extended into the nonlinear 
regime (Section~\ref{sec5}). We also provide independent 
numerical support in favour of our analytical 
findings on the dynamics~(Sections~\ref{sec5}~\&~\ref{sec6}). 

\section{The mathematical conditions of spherically 
symmetric accretion}
\label{sec2}
The mathematical problem that was set up by~\citet{bon52} himself 
and that is now taken up as a starting model in accretion-related 
texts~\citep{skc90,fkr02}, involves two coupled fields, the 
local flow velocity, $v$, and the local density, $\rho$, of the 
compressible accreting fluid. 
These two coupled fields are governed by the continuity equation,
\begin{equation}
\label{con}
\frac{\partial \rho}{\partial t} + \frac{1}{r^2} 
\frac{\partial}{\partial r}\left(\rho vr^2\right) =0,
\end{equation}
and the inviscid Euler equation,
\begin{equation}
\label{euler} 
\frac{\partial v}{\partial t} + v\frac{\partial v}{\partial r}
+ \frac{1}{\rho}\frac{\partial P}{\partial r} 
+ \Phi^\prime (r)=0,
\end{equation} 
tailored as they are, according to the requirements of spherical
symmetry. 
In the latter equation, the local pressure, $P$, is expressed 
in terms of $\rho$, by invoking a general polytropic 
prescription, $P=k\rho^\gamma$, in which $\gamma$, the polytropic 
exponent, varies over the range (limited by isothermal
and adiabatic conditions),  
$1 \leq \gamma \leq c_{\mathrm P}/c_{\mathrm V}$,
with $c_{\mathrm P}$ and $c_{\mathrm V}$ being the two 
coefficients of specific heat capacity of a gas~\citep{sc39}. 
The polytropic prescription is suited well 
for the study of open systems like astrophysical flows.
Making use of both $P$ and $\rho$, we scale the bulk 
flow velocity, $v$, in terms of a natural hydrodynamic
scale of speed, $c_{\mathrm s}$, which is the local speed 
of sound. This speed can be noted from 
$c_{\mathrm s}^2= \partial P/\partial \rho
= \gamma k\rho^{\gamma -1}$.  
The non-self-gravitating bulk flow is driven by the 
external gravity of a central accretor, whose 
potential is $\Phi(r)$. In equation~(\ref{euler}) the
driving force arising due to this potential is implied by 
its spatial derivative (represented by the prime). In the case of 
stellar accretion, the flow is driven by the Newtonian potential, 
$\Phi(r)=-GM/r$. In studies of 
accretion onto a non-rotating black hole, it is often expedient  
to dispense with the rigour of general relativity, and instead 
use a pseudo-Newtonian potential~\cite{pw80,abn96} 
to mimic the general relativistic effects of Schwarzschild 
space-time geometry in a Newtonian construct of space and 
time (see~\cite{ds01} and references therein). The choice 
of a pseudo-Newtonian potential, however, does 
not affect overmuch the general arguments regarding the 
stability of the flow.  

With the functions, $P$ and $\Phi(r)$, specified thus,  
equations~(\ref{con})~and~(\ref{euler}) give a complete
description of the hydrodynamic flow in terms of the two fields,
$v(r,t)$ and $\rho (r,t)$. The steady solutions of the flow 
are obtained from these dynamic variables by making explicit 
time-dependence disappear, i.e. 
$\partial v/ \partial t = \partial \rho /\partial t = 0$. The 
resulting differential equations, with full spatial 
derivatives only, can then be easily integrated to get the 
stationary global solutions of the flow~\citep{bon52,fkr02}.    
A noticeable feature of these stationary solutions is that 
they are invariant under the transformation
$v \longrightarrow -v$, i.e. the mathematical problems of 
inflows ($v<0$) and outflows ($v>0$) are identical in the 
steady state~\citep{arc}. This invariance has adverse 
implications for critical flows in accretion processes. 
Critical solutions pass through saddle points in the stationary 
phase portrait of the flow~\citep{rb02,rrbon07}, but generating 
a stationary solution through a saddle point will be impossible 
by any physical means, because it calls for an infinite precision 
in the required outer boundary condition~\citep{rb02}.
Nevertheless, criticality is not a matter of doubt in accretion
processes~\citep{bon52,gar79,shateu83}. The key
to resolving this paradox lies in considering explicit
time-dependence in the flow, because of which, as we note from
equations~(\ref{con})~and~(\ref{euler}), the invariance under
the transformation, $v \longrightarrow -v$, breaks down.
Obviously then, a choice of inflows $(v <0)$ or
outflows $(v >0)$ has to be made at the very beginning
(at $t=0$), and solutions generated thereafter
will be free of all the difficulties associated with the
presence of a saddle point in the stationary flow.

On imposing various boundary conditions on the steady
integral solutions, multiple classes of flow result~\citep{fkr02}. 
Of these, the one that attracts attention in accretion studies 
obeys the boundary conditions, $v \longrightarrow 0$ as
$r \longrightarrow {\infty}$ (the outer boundary condition) 
and $v> c_{\mathrm s}$ for small values of $r$. It is quite
obvious that this solution is transonic in nature, with its
bulk flow velocity overcoming the local speed of sound at a 
particular point in space, $r_{\mathrm c}$, the critical radius  
of the flow~\citep{skc90,fkr02,rb02}. For a flow driven 
simply by the Newtonian potential, there is only one such 
critical radius. 
With the choice of a pseudo-Newtonian potential, multiple 
values of $r_{\mathrm c}$ could result, but practically 
speaking there would be only one physically relevant critical 
point, through which an integral solution could pass and attain 
the transonic state~\citep{mrd07}.  
It was argued by~\citet{bon52} that among all the feasible 
stationary solutions by which a fluid element may reach the 
accretor, after having started under highly subsonic conditions 
on very large length scales, the actual trajectory chosen 
will be the one that is transonic in 
nature --- the~\citet{bon52} solution. This thought was 
guided by the criteria that with no restrictive inner boundary 
condition, the accretion rate will be as high as possible and 
the corresponding energy configuration of the flow shall be 
the lowest one~\citep{gar79}. In the stationary regime, 
the transonic solution does conform to these requirements. 
Under the 
approximation of a ``pressureless" motion of a fluid in a 
gravitational field~\citep{shu}, qualified support for 
transonicity also came later from a non-perturbative dynamic 
perspective~\citep{rb02,rrbon07}. 
No definitive conclusion about transonicity
can be drawn on the basis of a linearized perturbative
analysis~\citep{gar79}.   

So far as generating the transonic flow is concerned, 
the non-perturbative dynamic evolution 
of global $v(r,t)$ and $\rho (r,t)$ profiles 
is very crucial indeed. Certainly, all the feasible stationary 
inflow solutions obey the outer boundary conditions 
that are on large spatial scales, $v(r)\longrightarrow 0$ and 
$\rho (r) \longrightarrow \rho_\infty$, where $\rho_\infty$
is the constant ``ambient" value of the density field very far 
away from the accretor~\citep{fkr02}. It is the way in which 
the two fields evolve close to the accretor that determines 
if the transonic state would be achieved or not. 
The dynamic process should be conceived of ideally as one 
in which both the velocity and density fields, $v(r,t)$ and
$\rho (r,t)$, are uniform initially for all values of $r$,
in the absence of any driving force. Then with 
the introduction of a gravitational field (made effective 
at $t=0$), the hydrodynamic fields, $v$ and $\rho$, start 
evolving in time. If the temporal growth of $v$ outpaces 
the temporal growth of $\rho$ (to which $c_{\mathrm s}$ is 
connected) at small values of $r$, then the final stationary 
infall process will be transonic. Otherwise, the final 
stationary infall process will be globally
subsonic, with $v(r) \longrightarrow 0$
as $r \longrightarrow 0$~\citep{pso80}. Simple as this 
physical description may sound, it poses a formidable 
mathematical challenge, because the non-perturbative evolution 
of the velocity and density fields in spherically symmetric 
accretion entails working with a coupled set 
of nonlinear partial differential equations, as implied by
equations~(\ref{con})~and~(\ref{euler}). And where nonlinear 
equations are involved, we have to tread with caution, 
especially since no analytical solution of the dynamic 
problem exists for spherically symmetric accretion.   

\section{Nonlinearity in the perturbative analysis}
\label{sec3}
Equations~(\ref{con})~and~(\ref{euler}) are integrated in 
their stationary limits, and the resulting velocity and 
density fields have only spatial
profiles, $v \equiv v_0(r)$ and $\rho \equiv \rho_0(r)$. 
A standard practice in perturbative analysis~\citep{pso80} 
is to apply small time-dependent, radial perturbations
on the stationary profiles, $v_0(r)$ and $\rho_0(r)$, and 
then linearize the perturbed quantities. This, however, 
does not offer much insight into the time-dependent 
evolutionary aspects of the hydrodynamic flow. 
So the next logical step is to incorporate nonlinearity in 
the perturbative method. With the inclusion of nonlinearity 
in progressively higher orders, the perturbative analysis 
incrementally approaches the actual time-dependent evolution 
of global 
solutions, that starts with a given stationary profile 
at $t=0$ (to make physical sense, this initial profile 
is very much subsonic at all spatial points).

The prescription for the time-dependent radial perturbation 
is $v(r,t)=v_0(r)+v^{\prime}(r,t)$
and $\rho (r,t)=\rho_0(r)+\rho^{\prime}(r,t)$, in which the
primed quantities indicate a perturbation about a 
stationary background.  
We define a new variable, $f(r,t)=\rho vr^2$, following
a similar mathematical procedure employed by~\citet{pso80} 
and~\citet{td92}. This variable emerges
as a constant of the motion from the stationary limit of 
equation~(\ref{con}). This constant, $f_0$, can be identified
with the matter flow rate, within a geometrical 
factor of $4\pi$~\citep{fkr02},
and in terms of $v_0$ and $\rho_0$, it is given 
as $f_0 = \rho_0 v_0r^2$. On applying the perturbation 
scheme for $v$ and $\rho$, the perturbation in $f$,
without losing anything of nonlinearity, is derived as 
\begin{equation}
\label{pertef}
\frac{f^\prime}{f_0} = \frac{\rho^\prime}{\rho_0} 
+\frac{v^\prime}{v_0}+\frac{\rho^\prime}{\rho_0}
\frac{v^\prime}{v_0}.
\end{equation} 
The foregoing relation connects the perturbed quantities, 
$v^\prime$, $\rho^\prime$ and $f^\prime$, to one another. 
To get a relation between only $\rho^\prime$ and 
$f^\prime$, we have to go back to equation~(\ref{con}), 
and apply the perturbation scheme on it. This will result in
\begin{equation}
\label{pertrho}
\frac{\partial \rho^\prime}{\partial t}
= -\frac{1}{r^2} 
\frac{\partial f^\prime}{\partial r}.
\end{equation} 
To obtain a similar relationship solely between $v^\prime$ and
$f^\prime$, we combine the conditions 
given in equations~(\ref{pertef})~and~(\ref{pertrho}), to get
\begin{equation}
\label{pertvee}
\frac{\partial v^\prime}{\partial t} = \frac{v}{f}
\left(\frac{\partial f^\prime}{\partial t}+ 
v\frac{\partial f^\prime}{\partial r} \right).
\end{equation} 
%and its second partial time derivative, 
%\begin{equation}
%\label{pertvee2}
%\frac{\partial^2 v^\prime}{\partial t^2}=\frac{\partial}{\partial t}
%\left[\frac{v}{f}\left(\frac{\partial f^\prime}{\partial t}\right)
%\right] + \frac{\partial}{\partial t}
%\left[\frac{v^2}{f}\left(\frac{\partial f^\prime}{\partial r}\right)
%\right].
%\end{equation} 
In equations~(\ref{pertef}),~(\ref{pertrho})~and~(\ref{pertvee}), 
%\&~(\ref{pertvee2}), 
all orders of nonlinearity have been maintained. Adhering to the 
same principle, applying the perturbation scheme in 
equation~(\ref{euler}), and taking its second-order partial time
derivative, will yield
\begin{equation}
\label{perteuler}
\frac{\partial^2 v^\prime}{\partial t^2} + 
\frac{\partial}{\partial r}\left( 
v\frac{\partial v^\prime}{\partial t} + 
\frac{c_{\mathrm s}^2}{\rho}\frac{\partial \rho^\prime}
{\partial t} \right) =0.
\end{equation}   
In deriving this expression, all terms involving only  
the stationary flow have vanished due to 
taking a partial time derivative. This is slightly different
from the practice of extracting the stationary part of 
equation~(\ref{euler}) and making it disappear by setting 
its value as zero. Now making use of 
equations~(\ref{pertrho}),~(\ref{pertvee}) and 
the second partial time derivative of equation~(\ref{pertvee}), 
we obtain a fully nonlinear equation of the perturbation 
from equation~(\ref{perteuler}), in a symmetric form as
\begin{equation}
\label{perteq}
\frac{\partial}{\partial t}\left(h^{tt}\frac{\partial f^\prime}
{\partial t}\right)+
\frac{\partial}{\partial t}\left(h^{tr}\frac{\partial f^\prime}
{\partial r}\right) 
+\frac{\partial}{\partial r}\left(h^{rt}\frac{\partial f^\prime}
{\partial t}\right) +
\frac{\partial}{\partial r}\left(h^{rr}\frac{\partial f^\prime}
{\partial r}\right)=0,
\end{equation} 
in which, 
\begin{equation}
\label{aitch}
h^{tt}=\frac{v}{f},\,\,
h^{tr}=h^{rt}=\frac{v^2}{f},\,\,
h^{rr}=\frac{v}{f} \left(v^2 -c_{\mathrm s}^2\right).  
\end{equation} 
Going by the symmetry of equation~(\ref{perteq}), it can be recast
in a compact form as 
\begin{equation}
\label{compact}
\partial_\mu \left(h^{\mu \nu}\partial_\nu 
f^\prime \right) =0, 
\end{equation} 
with the Greek indices running from $0$ to $1$, under the 
equivalence that $0$ stands for $t$ and $1$ stands for $r$.  
Equation~(\ref{compact}), or equivalently, equation~(\ref{perteq}), 
is a nonlinear equation containing arbitrary orders of nonlinearity 
in the perturbative expansion. All of the nonlinearity is carried 
in the metric elements, $h^{\mu \nu}$, involving the exact field 
variables, $v$, $c_{\mathrm s}$ and $f$, as opposed to containing 
only their stationary background counterparts~\citep{vis98,su02}. 
This is going into the realm of nonlinearity, because $v$ and 
$c_{\mathrm s}$ depend on $f$, while $f$ is related to $f^\prime$ 
as $f=f_0+f^\prime$. If we were to have worked with a linearized 
equation, then $h^{\mu \nu}$, containing only the zeroth-order 
terms, can be read from the matrix, 
\begin{equation}
\label{matrix}
h^{\mu \nu }=\frac{v_0}{f_0}
\begin{pmatrix}
1 \hfill & v_0 \\
v_0 & v_0^2 - c_{\mathrm s0}^2 \hfill 
\end{pmatrix}, 
\end{equation}
in which $c_{\mathrm s0}(r)$ is the stationary value of the
local speed of sound. An implication of the foregoing matrix
is that under steady conditions, an acoustic disturbance in the
fluid propagates with the speed, $c_{\mathrm s0}$.

Now, in Lorentzian geometry the d'Alembertian of a scalar field in 
curved space is expressed in terms of the metric, $g_{\mu \nu}$, as
\begin{equation}
\label{alem}
\Delta \varphi \equiv \frac{1}{\sqrt{-g}}
\partial_\mu \left({\sqrt{-g}}\, g^{\mu \nu} 
\partial_\nu \varphi \right), 
\end{equation}
with $g^{\mu \nu}$ being the inverse of the matrix implied
by $g_{\mu \nu}$~\citep{vis98,blv11}. We look for an equivalence 
between $h^{\mu \nu }$ and $\sqrt{-g}\, g^{\mu \nu}$ by 
comparing equations~(\ref{compact})~and~(\ref{alem})
with each other, and we see that 
equation~(\ref{compact}) gives an expression of $f^{\prime}$ 
that is of the type shown by equation~(\ref{alem}). In the 
linear order, the metrical part of equation~(\ref{compact}), 
as equation~(\ref{matrix}) shows it, may then be extracted, 
and its inverse will indicate the existence of an acoustic 
horizon, when $v_0^2 = c_{\mathrm s0}^2$. In the case of a 
radially inflowing astrophysical fluid, this horizon is due 
to an acoustic black hole. The radius of the horizon is the 
critical radius of the flow, $r_{\mathrm c}$. It cannot be 
breached by an acoustic disturbance (carrying any information)
propagating against the bulk outflow, after having originated 
in the super-critical region, where 
$v_0^2 > c_{\mathrm s0}^2$ and $r < r_{\mathrm c}$.
We can thus say that the flow of information
across the acoustic horizon is uni-directional. We can also 
arrive at this very conclusion by considering spherically 
symmetric accretion as an irrotational, inviscid and barotropic 
fluid flow (a potential flow), whose velocity is 
the gradient of a scalar potential. Then we may impose a  
perturbation on this scalar potential~\citep{vis98,bil99,blv11}, 
but we stress that in contrast to this approach of exploiting 
the conservative nature of the flow to craft a scalar potential
and then perturbing it, 
the derivation of equation~(\ref{compact}) makes use of  
the continuity condition. Our claim is that   
the latter method is more robust because the continuity 
condition is based on matter conservation, which is a firmer
conservation principle than that of energy conservation
(especially where open astrophysical flows are concerned), 
on which the conventional scalar-potential approach is founded. 
Regardless of the approach chosen, however, what we realize is  
that the physics of supersonic acoustic
flows closely corresponds to many features of black hole physics. 
All infalling matter crosses the event horizon of a black hole 
maximally, i.e. at the greatest possible speed. By analogy the 
same thing may be said of matter crossing the sonic horizon 
in spherically symmetric inflows. We recall that a  
long-standing conjecture about spherically 
symmetric accretion onto a point sink is that the
transonic solution crosses the sonic horizon at the 
greatest possible rate~\citep{bon52,gar79}. That a perturbative 
treatment may hint at this conjecture is noteworthy, 
because conventional wisdom has it that perturbative techniques 
are inadequate on this point~\citep{gar79}.

The perspective of an analogue horizon is valid only as far as 
the linear ordering 
goes. When nonlinearity is to be accounted for, then instead 
of equation~(\ref{matrix}), it is equation~(\ref{aitch}) 
that defines the elements, $h^{\mu \nu}$, depending on 
the order of nonlinearity that we wish to retain (in principle
we could go up to any arbitrary order). The first serious 
consequence of including nonlinearity is that the notion 
of static and zero-order $h^{\mu \nu}$, as stated in 
equation~(\ref{matrix}), will have to be abandoned. 
This view conforms to a numerical study 
of~\citet{macmal08}, who, for spherically
symmetric accretion, showed that if the
perturbations were to become strong then acoustic
horizons would suffer a shift about their 
static position, and the analogy between an acoustic horizon
and the event horizon of a black hole would
appear limited. In short, we lose the argument in favour of 
the static transonic condition (a steady inflow solution
crossing the sonic horizon).
For all that, a most remarkable fact 
that has emerged in consequence of including nonlinearity in 
the perturbative analysis, is that notwithstanding the order 
of nonlinearity that we may adopt, the symmetric 
form of the metric equation will remain unchanged, as shown 
very clearly by equation~(\ref{compact}). Precedence of the 
survival of this symmetry under nonlinear conditions, can be 
found in the fluid problems of the hydraulic jump~\cite{rbpla07} 
and spherically symmetric outflows of nuclear matter~\cite{sbbr13}.   

\section{Standing waves on globally subsonic steady inflows}
\label{sec4}
All physically relevant inflow solutions obey the outer 
boundary condition, $v(r) \longrightarrow 0$ as 
$r \longrightarrow \infty$. In addition, if the solution
is globally subsonic, then the inner boundary condition is
$v(r) \longrightarrow 0$ as $r \longrightarrow 0$. From 
the point of view of a gravity-driven evolution of an inflow 
solution to a transonic state, the subsonic flows have 
great importance, because the initial state of an 
evolution, as well as the intermediate states in the 
march towards transonicity, should realistically be subsonic. 
So the stability of globally subsonic solutions must have 
a significant bearing on how a transonic solution will develop 
eventually. Imposing an Eulerian perturbation on subsonic 
inflows, their stability was studied by~\citet{pso80}, and 
the amplitude of the perturbation in this case maintained 
a constant profile in time, i.e. it was marginally stable. 
In this respect we may say that 
the solutions do not exhibit any obvious instability. 
However, it is never prudent to extend this argument 
too far, especially when we consider nonlinearity in 
the perturbative effects, as it rightly ought to be done 
in a fluid flow problem.  

Equation~(\ref{perteq}) gives a nonlinear equation of the 
perturbation, accommodating nonlinearity up to any desired order.
It is important to realize here that the derivation of 
equation~(\ref{perteq}) is pertinent to any kind of stationary 
background solution (transonic or subsonic), with the only 
restriction being that the perturbation is radial. Thereafter
its uses may vary, and here we apply this equation 
to study the stability 
of stationary subsonic flows in a nonlinear regime. 
Following the mathematical procedure of~\citet{pso80}, 
we design the perturbation to behave like a standing 
wave about a globally subsonic stationary solution, 
obeying the boundary condition that the spatial part 
of the perturbation vanishes at two radial points in 
the spherical geometry, one at a great distance from the 
accretor (the outer boundary), and the other very close to it 
(the inner boundary).  
We confine our mathematical treatment involving nonlinearity 
to the second order only (the lowest order
of nonlinearity). Even simplified so, the entire procedure
will still bear much of the complications associated with 
a nonlinear problem. The restriction of not going beyond
the second order of nonlinearity implies that all  
$h^{\mu \nu}$ in equation~(\ref{aitch}) will 
contain primed quantities in their first power only. 
Taken together with equation~(\ref{perteq}), this will
preserve all terms which are nonlinear in the second 
order. So, carrying out the necessary expansion of 
$v=v_0+v^\prime$, $\rho = \rho_0 +\rho^\prime$ and 
$f=f_0+f^\prime$ in equation~(\ref{aitch}), up to 
the first order only, and defining a new set of metric
elements, $q^{\mu \nu} = f_0 h^{\mu \nu}$, we obtain     
\begin{equation}
\label{compact2}
\partial_\mu \left(q^{\mu \nu}\partial_\nu 
f^\prime \right) =0, 
\end{equation}
in which $\mu$ and $\nu$ are 
to be read just as in equation~(\ref{compact}). 
In the preceding expression, the elements, $q^{\mu \nu}$,
carry all the three perturbed quantities, $\rho^\prime$, 
$v^\prime$ and $f^\prime$. The next process to perform 
is to substitute both $\rho^\prime$ and $v^\prime$ in 
terms of $f^\prime$, since equation~(\ref{compact2}) is 
over $f^\prime$ only. To make this substitution possible,
first we have to make use of equation~(\ref{pertef}) 
to represent $v^\prime$ in terms of $\rho^\prime$ and 
$f^\prime$ in all $q^{\mu \nu}$. While doing so, we ignore
the product term of $\rho^\prime$ and $v^\prime$ in 
equation~(\ref{pertef}), because 
including it will raise equation~(\ref{compact2}) to 
the third order of nonlinearity. Once $v^\prime$ has 
been substituted in this manner,
we have to write $\rho^\prime$ in terms of $f^\prime$. 
This can be done by invoking the linear relationship 
suggested by equation~(\ref{pertrho}), 
with the reasoning that if $\rho^\prime$ and $f^\prime$ 
are both multiplicatively separable functions of space 
and time, with an exponential time part (all of which 
are standard mathematical prescriptions in any analysis 
that requires working with standing waves), then
\begin{equation}
\label{rhoeflin}
\frac{\rho^\prime}{\rho_0} = \sigma (r) 
\frac{f^\prime}{f_0},
\end{equation}
with $\sigma$ being a function of $r$ only (which lends 
a crucial advantage 
in simplifying much of the calculations to follow). 
The exact functional form of $\sigma (r)$ is determined 
from the way the spatial part of $f^\prime$ is prescribed,  
but on general physical grounds it 
stands to reason that when $\rho^\prime$, 
$v^\prime$ and $f^\prime$ are all real fluctuations, $\sigma$ 
should likewise be real.\footnote{Treating the perturbation 
as a high-frequency travelling wave, it was shown 
by~\citet{pso80} that  
$\sigma (r)=v_0\left(v_0 \pm c_{\mathrm s0}\right)^{-1}$, 
when the spatial part of $f^\prime$ was cast as a power
series in the {\it Wentzel-Kramers-Brillouin} approximation.} 
Following all of these algebraic details, the
elements, $q^{\mu \nu}$, in equation~(\ref{compact2}), can
finally be expressed entirely in terms of $f^\prime$ as
\begin{equation} 
\label{que}
q^{tt}= v_0\left(1 +\epsilon \xi^{tt}
\frac{f^\prime}{f_0}\right),\,\,
q^{tr}= v_0^2\left(1 +\epsilon \xi^{tr} 
\frac{f^\prime}{f_0}\right),\,\,
q^{rt}= v_0^2\left(1 +\epsilon \xi^{rt} 
\frac{f^\prime}{f_0}\right),\,\,
q^{rr}= v_0\left(v_0^2-c_{\mathrm s0}^2\right)+\epsilon v_0^3
\xi^{rr} \frac{f^\prime}{f_0},
\end{equation} 
in all of which, $\epsilon$ has been introduced as a nonlinear
``switch" parameter to keep track of all the nonlinear terms. 
When $\epsilon =0$, only linearity remains, and in this 
limit we converge to the familiar result indicated  
by equation~(\ref{matrix}). In the opposite extreme, when  
$\epsilon =1$, in addition to the linear effects, the lowest 
order of nonlinearity (the second order) becomes activated
in equation~(\ref{compact2}), and the linearized stationary
conditions of an acoustic horizon get disturbed due to the 
nonlinear $\epsilon$-dependent terms. This feature was 
pointed out numerically by~\citet{macmal08}. 
Equation~(\ref{que}) also contains the factors, $\xi^{\mu \nu}$, 
all of which are to be read as
\begin{equation}
\label{zyees}
\xi^{tt}= -\sigma,\,\,
\xi^{tr}=\xi^{rt}=1-2\sigma,\,\,
\xi^{rr}= 2 - \sigma \left[3+(\gamma -2)
\frac{c_{\mathrm s0}^2}{v_0^2}\right].  
\end{equation}
Taking equations~(\ref{compact2}),~(\ref{que})~and~(\ref{zyees}) 
together, we finally obtain a nonlinear equation of the 
perturbation, completed up to the second order, without the 
loss of any relevant term.

To render equation~(\ref{compact2}), along with all $q^{\mu \nu}$
and $\xi^{\mu \nu}$, into a workable form, it will first have to
be written explicitly, and then divided throughout by $v_0$. 
While doing so, the symmetry afforded by 
$\xi^{tr} = \xi^{rt}$ is also to be exploited. 
The desirable form of the equation of the perturbation should
be such that its leading term would be a second-order partial
time derivative of $f^\prime$, with unity as its coefficient. 
To arrive at this form, an intermediate step will involve a
division by $1 + \epsilon \xi^{tt} (f^\prime/f_0)$, which, 
binomially, is the equivalent of a multiplication by 
$1 - \epsilon \xi^{tt} (f^\prime/f_0)$, with a truncation 
applied thereafter. This is dictated by the simple principle 
that to keep only the second-order nonlinear terms, it 
will suffice to retain just those terms that carry $\epsilon$ 
in its first power. The result of this entire exercise is
\begin{eqnarray}
\label{pertorder2}
\frac{\partial^2 f^\prime}{\partial t^2}
+ 2\frac{\partial}{\partial r}
\left(v_0\frac{\partial f^\prime}{\partial t}\right) 
+ \frac{1}{v_0}\frac{\partial}{\partial r} 
\left[v_0\left(v_0^2-c_{\mathrm s0}^2\right)
\frac{\partial f^\prime}{\partial r}\right]%\nonumber\\ 
&+& \frac{\epsilon}{f_0}{\Bigg\{} \xi^{tt}
\left(\frac{\partial f^\prime}{\partial t}\right)^2  
+ \frac{\partial}{\partial r}\left(\xi^{rt} v_0 
\frac{\partial f^{\prime 2}}{\partial t}\right)%\nonumber\\
- \frac{v_0}{2}\frac{\partial \xi^{rt}}{\partial r}
\frac{\partial f^{\prime 2}}{\partial t}%\nonumber\\
+ \frac{1}{2v_0}\frac{\partial}{\partial r}
\left(\xi^{rr}v_0^3
\frac{\partial f^{\prime 2}}{\partial r}\right)\nonumber\\ 
&-& 2 \xi^{tt}f^{\prime}\frac{\partial}{\partial r}
\left(v_0\frac{\partial f^\prime}{\partial t}\right)%\nonumber\\ 
- \frac{\xi^{tt}f^\prime}{v_0}\frac{\partial}{\partial r} 
\left[v_0\left(v_0^2-c_{\mathrm s0}^2\right)
\frac{\partial f^\prime}{\partial r}\right]
{\Bigg\}}=0, 
\end{eqnarray} 
in which, if we set $\epsilon =0$, then what remains 
is the linear equation discussed in detail 
by~\citet{pso80} and~\citet{td92}. We apply 
a solution, $f^\prime (r,t)=R(r)\phi (t)$, in 
equation~(\ref{pertorder2}), with $R$ being 
a real function~\cite{polza}. After this, we multiply the 
resulting expression throughout by $v_0R$, perform some 
algebraic simplifications by partial integrations, to 
finally get  
\begin{eqnarray}
\label{aarphi}
\ddot{\phi} v_0 R^2 &+& \dot{\phi} \frac{\mathrm d}{{\mathrm d}r} 
\left(v_0 R\right)^2 + \phi \left\{ \frac{\mathrm d}{{\mathrm d}r} 
\left[\frac{v_0}{2}
\left(v_0^2-c_{\mathrm s0}^2\right)\frac{{\mathrm d}R^2}{{\mathrm d}r}
\right] -v_0\left(v_0^2-c_{\mathrm s0}^2\right)
\left(\frac{{\mathrm d}R}{{\mathrm d}r}\right)^2 \right\}%\nonumber\\ 
+ \frac{\epsilon}{f_0} \Bigg{[} 
\dot{\phi}^2 \xi^{tt}v_0R^3 \nonumber\\
&+& \dot{\phi}{\phi} \left[ \frac{\mathrm d}{{\mathrm d}r}
\left(\xi^{rt}v_0^2R^3\right) + \xi^{rt}\frac{v_0^2}{3}
\frac{{\mathrm d}R^3}{{\mathrm d}r} -\xi^{tt}R
\frac{\mathrm d}{{\mathrm d}r}\left(v_0R\right)^2 \right]%\nonumber\\
+ \phi^2 \Bigg{\{} v_0\left(v_0^2-c_{\mathrm s0}^2\right)
\frac{{\mathrm d}R}{{\mathrm d}r}\frac{\mathrm d}{{\mathrm d}r} 
\left(\xi^{tt}R^2\right)\nonumber \\
&-&\xi^{rr}v_0^3R
\left(\frac{{\mathrm d}R}{{\mathrm d}r}\right)^2%\nonumber\\ 
- \frac{\mathrm d}{{\mathrm d}r}\left[\xi^{tt}\frac{v_0}{3}
\left(v_0^2-c_{\mathrm s0}^2\right)
\frac{{\mathrm d}R^3}{{\mathrm d}r}\right]%\nonumber\\
+ \frac{\mathrm d}{{\mathrm d}r}\left(\xi^{rr}\frac{v_0^3}{3}
\frac{{\mathrm d}R^3}{{\mathrm d}r}\right)
\Bigg{\}} \Bigg{]}=0, 
\end{eqnarray} 
in which the overdots indicate full derivatives in time. Quite 
evidently, equation~(\ref{aarphi}) is a second-order nonlinear
differential equation in both space and time.  
We integrate all spatial dependence out of 
equation~(\ref{aarphi}), and then study the nonlinear features
of the time-dependent part. The integration over the spatial 
part will necessitate invoking two boundary conditions, one at 
a small value of $r$ (close to the accretor), and the other when 
$r \longrightarrow \infty$ (very far from the accretor). At  
both of these boundary points, the perturbation will have a
vanishing amplitude in time. It was reasoned by~\citet{pso80}
that globally subsonic inflow solutions offer conditions for 
the fulfilment of the two required boundary conditions, and 
simultaneously maintain a continuity of the background solution
in the interim region. The boundary conditions will ensure that
all the ``surface" terms of the integrals 
in equation~(\ref{aarphi}) will vanish
(which explains the tedious mathematical exercise to extract 
several such ``surface" terms). So after carrying out the required 
integration on equation~(\ref{aarphi}), over the entire region
trapped between the two specified boundaries, all that will 
remain is the purely time-dependent part, having the form, 
\begin{equation}
\label{onlytime}
\ddot{\phi} + \epsilon \left({\mathcal A}\phi 
+ {\mathcal B}\dot{\phi}\right) \dot{\phi} +{\mathcal C}\phi
+ \epsilon {\mathcal D} \phi^2 =0,
\end{equation} 
in which the constants, $\mathcal A$, $\mathcal B$, $\mathcal C$
and $\mathcal D$, are to be read as 
\begin{eqnarray}
\label{constants}
{\mathcal A}&=&
\frac{1}{f_0}
\left(\int v_0R^2\,{\mathrm d}r\right)^{-1}
\int\left[\xi^{rt}\frac{v_0^2}{3}
\frac{{\mathrm d}R^3}{{\mathrm d}r} -\xi^{tt}R
\frac{\mathrm d}{{\mathrm d}r}\left(v_0R\right)^2 \right]\,
{\mathrm d}r, \nonumber\\ 
{\mathcal B}&=& 
\frac{1}{f_0}\left(\int v_0R^2\,{\mathrm d}r\right)^{-1}
\int \xi^{tt}v_0R^3\, {\mathrm d}r, \nonumber\\
{\mathcal C}&=&
-\left(\int v_0R^2\,{\mathrm d}r\right)^{-1}
\int v_0\left(v_0^2-c_{\mathrm s0}^2\right)
\left(\frac{{\mathrm d}R}{{\mathrm d}r}\right)^2\,
{\mathrm d}r, \nonumber\\
{\mathcal D}&=&
\frac{1}{f_0}
\left(\int v_0R^2\,{\mathrm d}r\right)^{-1}
\int \left[v_0\left(v_0^2-c_{\mathrm s0}^2\right)
\frac{{\mathrm d}R}{{\mathrm d}r}\frac{\mathrm d}{{\mathrm d}r} 
\left(\xi^{tt}R^2\right)-\xi^{rr}v_0^3R
\left(\frac{{\mathrm d}R}{{\mathrm d}r}\right)^2\right]\,
{\mathrm d}r,  
\end{eqnarray}
respectively. The form in which equation~(\ref{onlytime}) has 
been abstracted is that of a general Li\'enard 
system~\citep{stro,js99}. All the terms of 
equation~(\ref{onlytime}), which carry the parameter, $\epsilon$, 
have arisen in consequence of nonlinearity. When we set 
$\epsilon =0$, we readily regain the linear results 
presented by~\citet{pso80}. To go beyond linearity, 
and to appreciate the role of nonlinearity
in the perturbation, we now have to understand  
the Li\'enard system that equation~(\ref{onlytime})
has brought forth.  

\section{Equilibrium conditions in the Li\'enard system and 
their implications}
\label{sec5}
The general mathematical form of a Li\'enard system is that of a  
nonlinear oscillator equation, going as~\citep{stro,js99} 
\begin{equation}
\label{lienard}
\ddot{\phi}+\epsilon{\mathcal H}(\phi ,\dot{\phi})\dot{\phi}
+{\mathcal V}^\prime (\phi)=0,
\end{equation}
in which, $\mathcal H$ is a nonlinear damping coefficient
(retaining the parameter, $\epsilon$, alongside $\mathcal H$, 
attests to its nonlinearity), and $\mathcal V$ is 
the ``potential" of the system (with the prime on it indicating 
its derivative with respect to $\phi$). Looking at 
equation~(\ref{onlytime}), we realize that 
${\mathcal H}(\phi, \dot{\phi}) 
={\mathcal A}\phi +{\mathcal B}\dot{\phi}$ and 
${\mathcal V}(\phi) 
={\mathcal C}({\phi^2}/{2})+\epsilon {\mathcal D}
({\phi^3}/{3})$, with the constant coefficients, 
$\mathcal A$, $\mathcal B$, $\mathcal C$ and $\mathcal D$
having to be read from equations~(\ref{constants}). 
To investigate the properties of the equilibrium points resulting 
from equation~(\ref{lienard}), we need to decompose this 
second-order differential 
equation into a coupled first-order dynamical system. To that 
end, on introducing a new variable, $\psi$, equation~(\ref{lienard})
can be recast as~\citep{js99} 
\begin{eqnarray}
\label{coupdyna}
\dot{\phi}&=& \psi \nonumber\\
\dot{\psi}&=& -\epsilon \left({\mathcal A}\phi 
+{\mathcal B}\psi \right)\psi 
-\left({\mathcal C}\phi +\epsilon {\mathcal D}\phi^2\right). 
\end{eqnarray} 
Equilibrium conditions are established with  
$\dot{\phi}=\dot{\psi}=0$. For the dynamical system indicated
by equations~(\ref{coupdyna}), this will immediately lead to two
equilibrium points on the $\phi$--$\psi$ phase plane. This is 
how it should be, because, having accommodated nonlinearity 
(by turning on the nonlinearity ``switch", i.e. setting $\epsilon =1$)
up to the second order only, equations~(\ref{coupdyna}) will 
be quadratic in both $\phi$ and $\psi$, yielding two equilibrium
solutions. Labelling these equilibrium points with a $\star$ 
superscript, we see that $(\phi^\star,\psi^\star)=(0,0)$ 
in one case, whereas in the other case, 
$(\phi^\star,\psi^\star)=(-{\mathcal C}/(\epsilon {\mathcal D}),0)$. 
So, one of the equilibrium points is located at the origin of 
the $\phi$--$\psi$ phase plane, while the location of the other 
will depend on both the sign and the magnitude of
${\mathcal C}/{\mathcal D}$. In effect, both the equilibrium 
points lie on the line, $\psi =0$, and correspond to the turning 
points of ${\mathcal V}(\phi)$. Higher orders of nonlinearity will 
simply proliferate equilibrium points on the line, $\psi =0$. 

\begin{figure}[floatfix]
\begin{center}
\includegraphics[scale=1.0, angle=0]{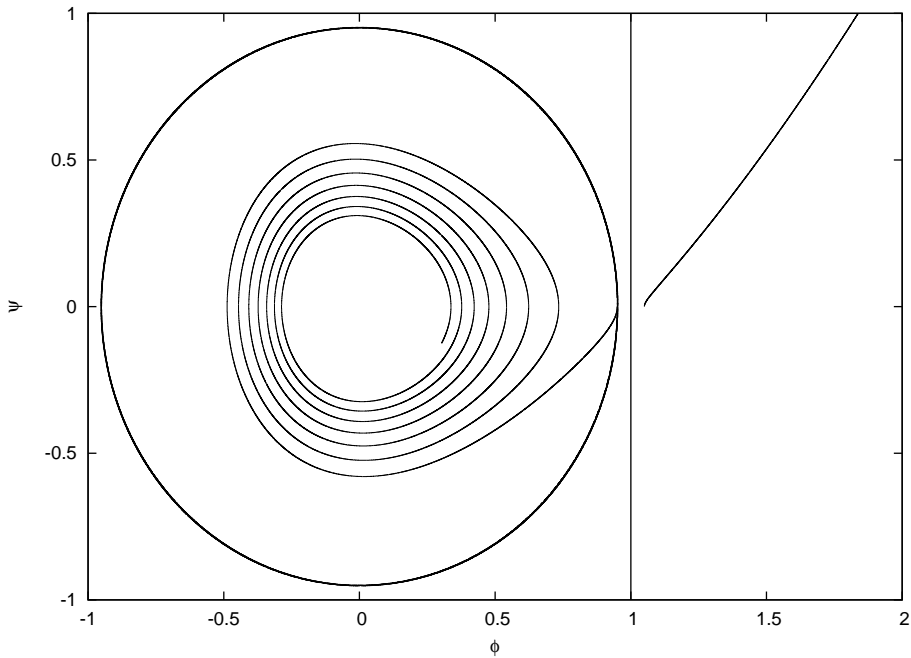}
\caption{\label{f1}\small{With a numerical integration of
equation~(\ref{lienard}) under chosen initial conditions,
three separate phase solutions are
plotted in the $\phi$--$\psi$ phase plane.
%for chosen values of $\mathcal A$, $\mathcal B$, $\mathcal C$ 
%and $\mathcal D$. 
The closed
elliptical solution corresponds to the case of $\epsilon =0$,
with ${\mathcal C}=1$. This is the phase solution representing
the linear perturbation on standing waves, with a centre-type
fixed point at the origin, $(0,0)$. The initial $(\phi, \psi)$
coordinates for tracing this trajectory in the
phase plane are $(0.95,0)$.
Retaining the same values of $\mathcal C$ and the initial
condition, the spiralling solution within the
elliptical envelope is obtained for $\epsilon =1$,
${\mathcal A}={\mathcal B}=0.03$ and ${\mathcal D}=-1$.
This solution depicts the phase-plane behaviour of the
second-order nonlinear perturbation.
With ${\mathcal C}=1$ and ${\mathcal D}=-1$, the coordinate
of the second fixed point (a saddle point) is set
at $(1,0)$. As long as the nonlinear perturbation starts
with a value
of $\phi <1$ (in this case its initial value is $0.95$),
it will always remain close to the linear regime and
stability can be maintained. This stability is evident
from the way the phase solution of the nonlinear perturbation
spirals towards the centre-type fixed point (which acts like
an attractor). A generalization of this argument is that
stability shall be achieved if
$\phi <\vert {\mathcal C}/{\mathcal D}\vert$, and the
values of $\mathcal A$ and $\mathcal B$ (whatever they
may be) will simply determine the rate at which the
nonlinear perturbation shall converge towards $(0,0)$.
A strong growth of the nonlinear
perturbation shall occur, once its initial value exceeds the
critical value of $\phi = \vert {\mathcal C}/{\mathcal D}\vert$.
This critical condition is indicated by the vertical line,
$\phi =1$, near the middle of the plot. To the left of this
line is the zone of stability, and to its right is the
zone of instability. Depending on the sign
of ${\mathcal C}/{\mathcal D}$,
the zone of instability will swivel either to the left
or to the right of the ellipse.
Setting the initial condition of the perturbation slightly
to the right of $\phi =1$, at $(1.05,0)$, the growth of the
perturbation is plainly visible, with an open trajectory
diverging outwards. For this diverging solution
the values of $\epsilon$,
${\mathcal A}$, ${\mathcal B}$, ${\mathcal C}$ and
${\mathcal D}$ are the same as they are for the spiralling
solution to the left of $\phi =1$.}}
\end{center}
\end{figure}

Having identified the position of the two equilibriums points, 
we now have to examine their stability. So we subject both 
equilibrium points to small perturbations, and then carry out
a linear stability analysis. The perturbation schemes on both 
$\phi$ and $\psi$ are
$\phi =\phi^\star +\delta \phi$ and $\psi =\psi^\star +\delta \psi$,
respectively. 
Applying these schemes on equations~(\ref{coupdyna}), and then 
linearizing in $\delta \phi$ and $\delta \psi$, will lead to 
the coupled linear dynamical system,  
\begin{eqnarray}
\label{lindyn}
\frac{\mathrm d}{{\mathrm d}t}\left(\delta \phi\right)&=& 
\delta \psi \nonumber\\
\frac{\mathrm d}{{\mathrm d}t}\left(\delta \psi\right)&=&
-{\mathcal V}^{\prime \prime}(\phi^\star)\delta\phi 
-\epsilon{\mathcal H}(\phi^\star,\psi^\star)\delta\psi,
\end{eqnarray} 
in which ${\mathcal V}^{\prime \prime}(\phi^\star)
= {\mathcal C} + \epsilon 2{\mathcal D} \phi^\star$. 
Using solutions of the type, $\delta \phi \sim \exp (\omega t)$
and $\delta \psi \sim \exp (\omega t)$, in equations~(\ref{lindyn}), 
the eigenvalues of the Jacobian matrix of the dynamical system
follow as 
\begin{equation}
\label{eigen}
\omega = -\epsilon
\frac{\mathcal H}{2} \pm
\sqrt{\epsilon^2
\frac{\mathcal H^2}{4} 
- {\mathcal V}^{\prime \prime}(\phi^\star)},
\end{equation}  
with $\mathcal H \equiv 
{\mathcal H}(\phi^\star,\psi^\star)$ having to be evaluated at 
the equilibrium points. 
Once the eigenvalues have been determined, it is now a simple 
task to classify the stability of an equilibrium point by 
putting its coordinates in equation~(\ref{eigen}). The equilibrium
point at the origin has the coordinates, $(0,0)$. Using these
coordinates in equation~(\ref{eigen}), the two roots of the 
eigenvalues are obtained as $\omega = \pm i \sqrt{\mathcal C}$. 
If $\mathcal C >0$, then the eigenvalues will be 
purely imaginary quantities, and consequently, the equilibrium 
point at the origin of the $\phi$--$\psi$ plane will be a 
centre-type point~\citep{js99}. And indeed, when the
stationary spherically symmetric inflow solution, about 
which the perturbation is constrained to behave like a 
standing wave, is globally subsonic, then $\mathcal C >0$, 
because in this situation, 
$v_0^2 < c_{\mathrm s0}^2$~\citep{pso80}. Viewed in the 
$\phi$--$\psi$ phase plane, the stationary solutions about this 
centre-type fixed point at the origin, $(0,0)$, will look like 
closed elliptical trajectories. 
This is identical to the phase portrait of a simple
harmonic oscillator with conserved total energy, and more to the
point, these solutions correspond entirely to the solutions with
unchanging amplitudes obtained by~\citet{pso80} in their linear 
stability analysis of standing waves on subsonic flows. Thus, 
in a linear framework, a marginal sense of stability is 
insinuated 
by the centre-type equilibrium point at the origin of the phase 
plane, because solutions in its neighbourhood are purely 
oscillatory in time, with no change in their amplitudes. 
While this conclusion was drawn 
by~\citet{pso80} in their linearized analysis of the standing 
wave, it could be arrived at equally correctly by   
setting $\epsilon =0$ (the linear condition) in 
equation~(\ref{eigen}). An illustration of this special
case is provided in Figure~\ref{f1}, which traces three
phase solutions of the Li\'enard system.
One of the solutions in this plot, obtained
for $\epsilon =0$ and corresponding physically to the linear
solution, is the closed elliptical trajectory about the centre-type
fixed point at $(0,0)$.

From dynamical
systems theory, centre-type points are known to be ``borderline"
cases~\citep{stro,js99}. In such situations, the linearized
treatment will show marginally stable behaviour, but robust 
stability or an instability may emerge immediately on accounting 
for nonlinearity~\citep{stro,js99}. This can be explained by a
simple but generic argument. Close to the coordinate, $(0,0)$,
equations~(\ref{coupdyna}) can be approximated in the linear
form as ${\dot \phi} = \psi$ and
${\dot \psi} \simeq -{\mathcal C}\phi$, which, of course,
gives a centre-type point, just like a simple harmonic oscillator.
Going further and accounting for the higher-order nonlinear terms,
equations~(\ref{coupdyna}) can be viewed as a coupled dynamical
system in the form,
${\dot \phi}={\mathcal F}(\phi,\psi)$ and
${\dot \psi}={\mathcal G}(\phi,\psi)$. Such a
system is said to be ``reversible" if
${\mathcal F}(\phi,-\psi)=-{\mathcal F}(\phi,\psi)$ and
${\mathcal G}(\phi,-\psi)={\mathcal G}(\phi,\psi)$, i.e. 
if $\mathcal F$ (or ${\dot \phi}$) is an odd function of 
$\psi$, and $\mathcal G$ (or ${\dot \psi}$) is an even function
of $\psi$~\citep{stro}. Centre-type points are robust under
this reversibility requirement. Now, a look at
equations~(\ref{coupdyna}) immediately reveals that $\dot{\psi}$
is not an even function of $\psi$. Therefore, the centre-type
point obtained due to a linearized analysis of
equations~(\ref{coupdyna}), is a fragile one. Ample evidence
of this feature can be found in the behaviour of the spiralling
solution (corresponding to the nonlinear case) in Figure~\ref{f1}.

The centre-type point at the origin of the phase plane has 
confirmed the known linear results. 
However, all of that is the most that a simple 
linear stability analysis
can bring forth. Accounting for nonlinearity,  
to its lowest order,   
another equilibrium point is obtained, in addition 
to the centre-type equilibrium point obtained by~\citet{pso80}. 
This second equilibrium point is an outcome of the quadratic
order of nonlinearity in the standing wave, and its
coordinates in the phase plane are
$(-{\mathcal C}/(\epsilon {\mathcal D}),0)$. Using these
coordinates in equation~(\ref{eigen}), the eigenvalues become
\begin{equation}
\label{eigenep2}
\omega = \frac{\mathcal{AC}}{2\mathcal{D}} \pm 
\sqrt{\left(\frac{\mathcal{AC}}{2\mathcal{D}}\right)^2 
+ {\mathcal C}}. 
\end{equation}
Noting as before, that $\mathcal C>0$, and that $\mathcal A$,
$\mathcal C$ and $\mathcal D$ are all real quantities, the 
inescapable conclusion is that the eigenvalues, $\omega$, are 
real quantities, with opposite signs. In other words, the second 
equilibrium point is a saddle point~\citep{js99}.
The position of this equilibrium point is at the
coordinate $(-{\mathcal C}/{\mathcal D},0)$ in the $\phi$--$\psi$
phase portrait. The absolute value of the abscissa of this
coordinate, $\vert {\mathcal C}/{\mathcal D} \vert$, represents
a critical threshold for the initial amplitude of the
perturbation. If this amplitude is less
than $\vert {\mathcal C}/{\mathcal D} \vert$,
then the perturbation will hover close to the linearized states
about the centre-type point, and stability shall
prevail. The spiralling solution in Figure~\ref{f1} gives a
clear demonstration of this fact. If, however, the amplitude
of the perturbation exceeds the critical value, i.e. if
$\vert \phi \vert > \vert {\mathcal C}/{\mathcal D} \vert$,
then we enter the nonlinear regime, and in time the perturbation
will undergo a divergence in one of its modes (for which
$\omega$ has a positive root). This state of affairs has
been depicted in the right side of the plot in Figure~\ref{f1},
showing a continuously diverging phase solution.  
Given that the eigenvalues,
$\omega$, have been yielded on using solutions of the type,
$\exp (\omega t)$, the e-folding time scale of
this growing mode of the perturbation is $\omega^{-1}$, with
$\omega$ having to be read from equation~(\ref{eigenep2}).

So, in the nonlinear regime, the simple fact that emerges is that
stationary subsonic global background solutions will become 
unstable under the influence of the perturbation. In the vicinity 
of a saddle point, if the initial amplitude of the perturbation 
is greater than $\vert {\mathcal C}/{\mathcal D} \vert$, then
the solutions will continue to diverge, and higher orders 
of nonlinearity (starting with the third order in this case) 
will not smother this effect~\citep{stro,js99}. Since a saddle 
point cannot be eliminated by the inclusion of higher orders of
nonlinearity~\citep{js99}, the best that we may hope for is that 
the instability may grow in time till it reaches a saturation 
level imposed by the higher nonlinear orders (but the instability 
will never be decayed down). This type of instability has a 
precedence in the laboratory fluid problem of the hydraulic 
jump~\citep{vol06,rbpla07}. 
While the discussion so far holds forth on the perturbative 
perspective, its crux lies in the far-reaching implications 
of the saddle point for the non-perturbative evolutionary 
dynamics. There exists no
analytical prescription for a full-blown time evolution of
nonlinear fluid equations. The next best thing in that case,
is to get as close to the true dynamics as possible, by the 
inclusion of progressively higher orders of nonlinearity in 
the perturbative treatment. It is evident that there can
be no transonic solution without gravity driving the infall
process. So, from a dynamic point of view, gravity starts 
the evolution towards the transonic state from an initial 
(and arguably nearly uniform) subsonic state, far away from
the critical conditions for transonicity. If, however, in the
real-time dynamics, the subsonic states are to encounter an 
instability that is attendant on a saddle point, then that 
should have adverse consequences for attaining a stable and 
stationary transonic end (the~\citet{bon52} solution)
through the dynamics.
 
\section{Dynamic evolution and instability}
\label{sec6}

\begin{figure}
\begin{center}
\includegraphics[scale=1.0, angle=0.0]{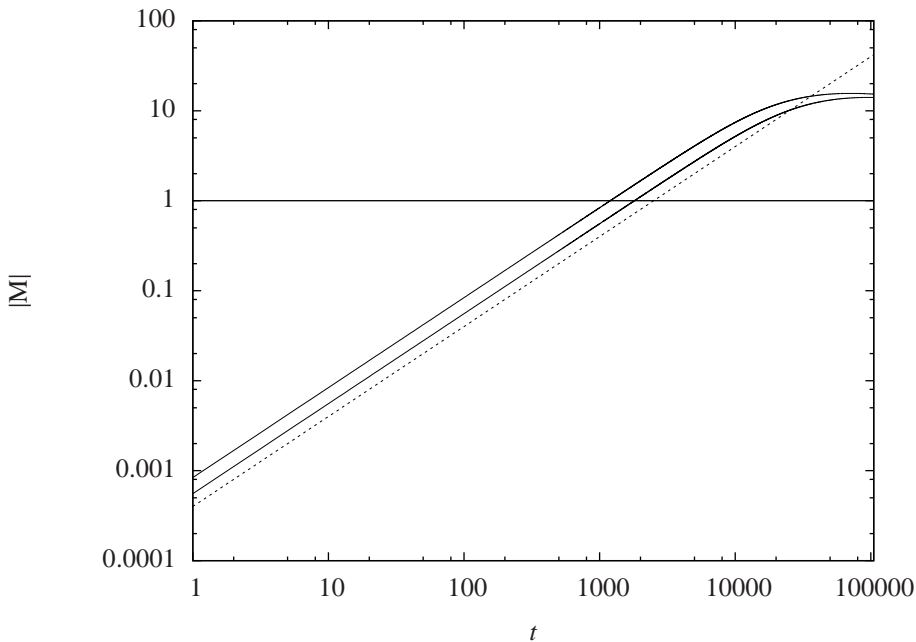}
\caption{\label{f2}\small{The time-dependent velocity field 
of a spherically symmetric accretion system is scaled as the 
Mach number, ${\mathrm M}$, and its time evolution is followed
at $0.9 r_{\mathrm c}$ (the upper full line) and 
$1.1 r_{\mathrm c}$ (the lower full line). In the early stages 
of the evolution, both lines indicate a power-law dynamics, 
$\vert {\mathrm M} \vert \sim t$. The lower full line,
pertaining to the dynamics in the subcritical region outside the
sonic radius, overshoots the expected steady subsonic limit of 
the~\citet{bon52} solution, and indicates the attainment of 
highly supersonic speeds, a fact that is clear from the way
the solution crosses the sonic barrier at
$\vert {\mathrm M} \vert =1$ (marked by the horizontal line 
in the plot). The 
lowermost dotted line is due to a toy function, used to test 
the correct exponent of the power law in the early stages of 
the evolution of $\vert {\mathrm M} \vert$. On later time 
scales the two full lines fall short of this power-law 
growth and converge towards an unexpectedly large
limiting value of $\vert {\mathrm M} \vert \sim 10$.}}
\end{center}
\end{figure}
A numerical exercise bears out the contention about how adversely
the saddle point affects the dynamics. The $\log$--$\log$ 
graph shown in Figure~\ref{f2} has been obtained by
numerically integrating equations~(\ref{con})~and~(\ref{euler})
by the finite-differencing technique. The integration has been 
performed under the flat initial conditions, $v=0$ and
$\rho= 10^{-21} {\mathrm{kg\, m^{-3}}}$ at $t=0$, for all 
values of $r$. Insofar as the purpose of the numerical
study is to know the effects of nonlinearity in the dynamic
evolution (not merely the perturbation), this initial condition
is the most appropriate one. To appreciate this 
fact, we recall that the perturbation scheme 
on the velocity field is $v(r,t) = v_0(r)+v^\prime (r,t)$. 
Now, linearity holds under the requirement, 
$\vert v^\prime /v_0 \vert \ll 1$.
Once the initial velocity field is prescribed to be 
zero everywhere, i.e. $v(r,0) =0$, then from a perturbative
perspective, this null initial condition can also be viewed
as a global background state, $v_0(r)=0$. Thereafter, any 
arbitrarily small perturbation, $v^\prime$, on this zero 
initial state, will be of a fully nonlinear order,
i.e. $v(r,t)=v^\prime (r,t)$.
The growth of such a small time-dependent perturbation
will become the actual dynamics of the global solutions, 
and this dynamics will be completely nonlinear.

In the numerics, the spherically symmetric flow is made to 
be driven by the Newtonian 
gravity of a star of mass, $M_\odot$. The polytropic index
of the accreting gas is given as $n=1.51$, and the ``ambient" 
conditions of the accreting gas are given by 
$c_{\mathrm s}(\infty)=10\,{\mathrm {km\, s^{-1}}}$ and 
$\rho_\infty = 10^{-21} {\mathrm{kg\, m^{-3}}}$. In terms of all 
these fixed parameters of the flow, the sonic radius is scaled as 
$r_{\mathrm c}=(n-1.5)
GM_\odot/2nc_{\mathrm s}^2(\infty)$~\citep{skc90,fkr02}.
The velocity field is scaled as the Mach number, ${\mathrm M}$, 
which is effectively the bulk velocity measured locally in terms 
of the speed of sound. In accretion flows, the bulk velocity
is conventionally assigned a negative sign~\cite{fkr02}, and 
so we present the absolute value of the Mach number, 
$\vert {\mathrm M} \vert$, in Figure~\ref{f2}. The time evolution 
of $\vert {\mathrm M} \vert$ is observed at $0.9 r_{\mathrm c}$ 
(inside the sonic radius) and $1.1 r_{\mathrm c}$ (outside
the sonic radius). The two full lines in Figure~\ref{f2}
suggest a linear growth on early time scales for the Mach number, 
i.e. $\vert {\mathrm M} \vert \sim t$, a feature that prevails 
over four orders of magnitude. On later time scales, both
lines show a convergence of the velocity field towards a limiting
value (${\mathrm M} \sim 10$), but this value is far greater 
than what is expected from the stationary~\citet{bon52}  
profile of the velocity field (which, in the neighbourhood of 
the sonic radius, is ${\mathrm M} \simeq 1$). This feature is 
particularly curious for the lower full line (plotted for 
$r > r_{\mathrm c}$), which shows  
an attainment of supersonic velocity, when even for the 
transonic solution, the steady velocity field outside the sonic 
radius has to be subsonic. This aspect of the dynamics is  
connected to the instability that arises due to the 
saddle point in the Li\'enard system, and near the steady 
sonic radius, also shows that nonlinearly strong signals may 
overshoot the constraints imposed by the acoustic 
horizon~\citep{macmal08}. The ultimate convergence towards an 
unexpectedly large limiting value (${\mathrm M} \sim 10$) 
occurs because of the saturating effect of the orders of 
nonlinearity higher than the second. Studies of the hydraulic 
jump phenomenon have reported similar saturation of a nonlinearly 
growing instability in the vicinity of the critical point of 
the flow~\cite{vol06,rbpla07}.  

\begin{figure}
\begin{center}
\includegraphics[scale=1.0, angle=0.0]{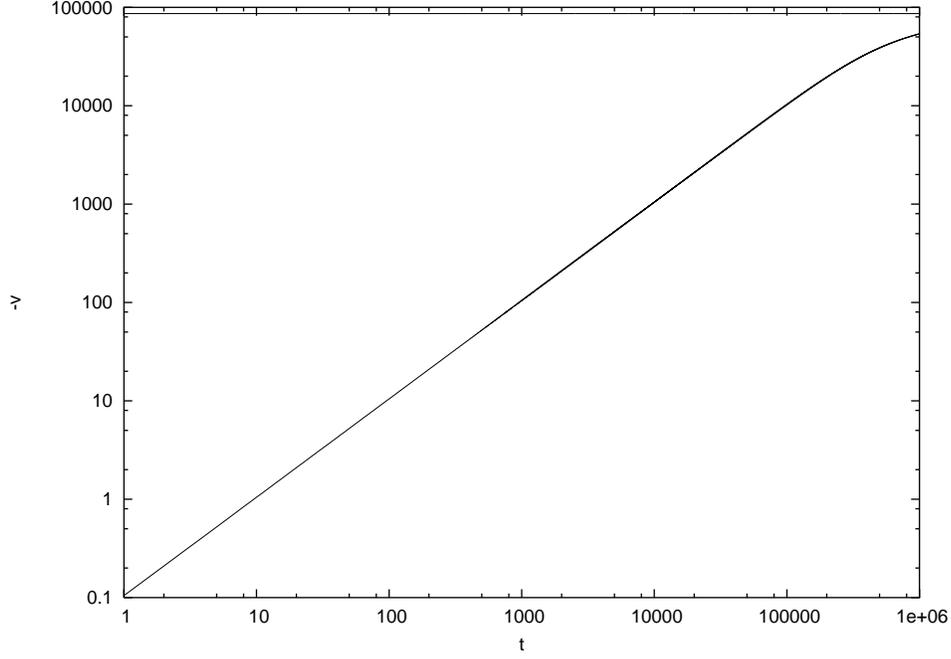}
\caption{\label{f3}\small{Under pressureless conditions in the
inflow ($v<0$), imposed on equation~(\ref{euler}),
the slope of this logarithmic plot shows that in the early stages
of the evolution, $-v$ varies linearly with $t$. Deviation from
linear growth sets in later. The horizontal line at the top
shows the free-fall value of the velocity field,
$\sqrt{2G{M_\odot}/51{r_\odot}}$, at a fixed radius of
$51r_\odot$. There is a convergence in the
dynamics towards this limiting value, and this may be seen in
contrast to the overshoot of the expected limit shown
in Figure~\ref{f2}.}}
\end{center}
\end{figure}
The purport of Figure~\ref{f2} can be compared with the convergence
exhibited by the time-evolving velocity field in a ``pressureless" 
fluid approximation~\citep{shu} towards the free-fall limiting
velocity of $\sqrt{2G{M_\odot}/r}$, 
with $r=51r_\odot$~\cite{rrbon07}. This convergence is shown 
graphically in Figure~\ref{f3}, and should be seen alongside 
Figure~\ref{f2} for contrast. The numerical convergence of the 
velocity, demonstrated by finite-differencing integration, is 
also in full 
agreement with the analytical treatment of the time evolution 
of the pressureless flow, carried out by the method of 
characteristics~\citep{rb02,rrbon07}. Equation~(\ref{euler}), 
rendered simple by setting $P=0$, under the requirement of 
a pressureless field driven by Newtonian gravity, appears as
\begin{equation}
\label{presfree} 
\frac{\partial v}{\partial t}+v\frac{\partial v}{\partial r} 
+ \frac{GM}{r^2}=0,
\end{equation}
and can be solved by the method of characteristics~\citep{ld97}.
The characteristic solutions are obtained from
\begin{equation}
\label{charcur} 
\frac{{\mathrm d}t}{1}= \frac{{\mathrm d}r}{v} 
= \frac{{\mathrm d}v}{-GM/r^2}. 
\end{equation}
First, solving the ${\mathrm d}v/{\mathrm d}r$ equation will give
\begin{equation}
\label{dvdrchar} 
\frac{v^2}{2}-\frac{GM}{r}=\frac{c^2}{2},  
\end{equation}
in which $c$ is an integration constant, coming from the 
spatial part of the characteristic equation. We use this result 
to solve the ${\mathrm d}r/{\mathrm d}t$ equation in 
equation~(\ref{charcur}), to get 
\begin{equation}
\label{drdt} 
\frac{2vr}{c r_{\mathrm s}} -\ln r -
\ln \left(\frac{v}{c}+1 \right)^2
- \frac{2ct}{r_{\mathrm s}} = {\tilde c},   
\end{equation}
with $\tilde{c}$ being another integration constant, and 
$r_{\mathrm s}$ being a length scale in the system, defined 
as $r_{\mathrm s}=2GM/c^2$. The general solution of 
equation~(\ref{charcur}) is given by the condition, 
$F({\tilde c})= c^2/2$, with $F$ being an arbitrary function, 
whose form is to be determined from the initial condition.
So, making use of 
equations~(\ref{dvdrchar}) and~(\ref{drdt}), we first write 
the general solution as
\begin{equation}
\label{gensol2} 
\frac{v^2}{2}-\frac{GM}{r}=F\left[\frac{2vr}{c{r_{\mathrm s}}}-\ln r 
-\ln \left(\frac{v}{c}+1\right)^2-\frac{2ct}{r_{\mathrm s}}\right], 
\end{equation}
to determine whose particular form, we then use the initial 
condition, $v=0$ at $t=0$ for all $r$. This gives us 
\begin{equation}
\label{partform} 
\frac{v^2}{2}-\frac{GM}{r}= -\frac{GM}{r}\left(\frac{v}{c}+1 
\right)^{-2}\exp\left(\frac{2vr}{cr_{\mathrm s}} 
-\frac{2ct}{r_{\mathrm s}}\right),  
\end{equation}
from which we see that for $t \longrightarrow \infty$, the 
right hand side of equation~(\ref{partform}) vanishes, and 
its left hand side implies a convergence of the velocity field 
towards its stationary free-fall limit, i.e. $v=\sqrt{2GM/r}$. 
Corresponding to the given initial condition, this is evidently
the final stationary solution associated with the lowest 
possible total energy, and the temporal evolution selects 
this solution. To envisage the evolution,  
the pressureless system, with a uniform velocity of 
$v=0$ everywhere, suddenly has a gravity mechanism activated 
in its midst at $t=0$. This induces a potential, $-GM/r$, at 
all points in space. The system then starts evolving to restore 
itself to another stationary state, with the velocity increasing 
according to equation~(\ref{partform}), so that 
for $t \longrightarrow \infty$, the total energy at all
points, $(v^2/2)-(GM/r)=0$, remains the same as what it 
was at $t=0$. This selection mechanism is compatible with 
the~\citet{bon52} criterion of a particular solution being 
chosen by the virtue of possessing the minimum energy. 

So how is the expected convergence achieved in the pressureless 
fluid approximation, while there seems to be an overshoot 
of the~\citet{bon52} limit by a very wide margin, when we
accommodate the pressure effects in the momentum balance condition
of the flow? Understanding of this question requires returning 
to the seminal work of~\citet{bon52}, in which he referred to some 
earlier works which had adopted the pressureless prescription, 
and had 
studied only the dynamical effects. \citet{bon52} himself adopted 
the opposite extreme of negligible dynamical effects but full pressure 
effects, as described by stationary flow equations. The difficulty 
arises when the dynamically evolving velocity field is nonlinearly
coupled to the dynamically evolving density field, through the 
pressure effects and the continuity equation, as can be seen 
in equations~(\ref{con})~and~(\ref{euler}). We contend 
that the instability shown by the Li\'enard system  
is connected to this nonlinear coupling of the two dynamic fields. 
More to the point, there will be no such Li\'enard system under 
the pressureless condition.
Not to mention also that the stationary
problem is not equipped to capture this nonlinear instability, 
and time-dependent linearization around stationary states is 
nearly as inadequate. 
Another interesting point in the complete dynamic-plus-pressure 
scenario is that the overshooting of the steady subsonic limit 
by the lower full line in Figure~\ref{f2} becomes more pronounced 
as the polytropic index, $n$, gets closer to the 
adiabatic limit of $n=1.5$ (or equivalently, $\gamma = 5/3$). 
Evidently, near this limit, the density field, $\rho$, 
fortified by a high power law, due to the polytropic          
prescription, $P=k\rho^\gamma$, robustly 
affects the dynamics of the momentum balance condition. 
So, coupling with the density field does serve to 
destabilize the velocity field, as it evolves dynamically. 

A final important point bears discussion. The two coupled 
fields, $v(r,t)$ and $\rho (r,t)$, are described by 
equations~(\ref{con})~and~(\ref{euler}), which are both 
first-order differential equations. This two-variable 
mathematical problem was reduced to a one-variable
problem, by introducing a new variable, $f=\rho vr^2$. 
The entire perturbative study and all consequent
analytical results, involve this new variable only, with the
perturbed quantity on its stationary background being
$f^\prime (r,t)$. The price of this one-variable convenience 
is a second-order differential equation, as can be seen from 
equation~(\ref{perteq}). While the analytical treatment 
in this work is based on the variable, 
$f(r,t)$, the numerics, however, has followed the growth of
the perturbation by means of the field, $v(r,t)$ (scaled
as the Mach number, as Figure~\ref{f2} indicates). In spite 
of this apparent difference, there is actually 
no contradiction between the analytical methods 
and the numerics, because from equation~(\ref{pertef}),
it is clear that 
$\vert v^\prime/v_0\vert \sim \vert f^\prime/f_0 \vert$, 
to a linear order. A precise analysis, of the kind leading
to equation~(\ref{rhoeflin}), yields
$\vert v^\prime/v_0 \vert =(1-\sigma)\vert f^\prime/f_0 \vert$.
This linear scaling between $f^\prime$ and $v^\prime$ suggests 
that the unstable growth behaviour of the one will be faithfully
captured by the other. We have chosen the variable, $v$ (scaled 
by the speed of sound), to numerically track the growth of the 
perturbation, because in accretion fluid mechanics, flow solutions 
are categorized in terms of how the bulk flow velocity measures 
up to the local speed of sound (e.g., transonic, subsonic, globally 
supersonic, etc.)

\section{Concluding remarks}
\label{sec7}
It will be well in order now to make some general remarks
about our work, to put it in perspective. First, 
accretion being a fluid problem, is very much within the realm 
of nonlinear dynamics. Our work 
addresses the nonlinear aspects of the dynamics of an accretion
process, by having recourse to the usual analytical tools of
nonlinear dynamics. One salient outcome of the nonlinear approach 
is obtaining an acoustic metric, in spite of accommodating 
nonlinearity completely. This marks a significant departure 
from the linear treatment (small perturbation) of the problem. 
Another new result of this work is the discovery of a
Li\'enard system (a nonlinear oscillator) in a very common 
and basic model of accretion. A noteworthy aspect of all 
these new results is that they have been extracted from a 
system that has been known to the astrophysical community for 
more than sixty years --- a conservative, non-self-gravitating, 
compressible fluid inflow, driven by Newtonian-like external 
gravitational fields, with coupled density and velocity fields. 
It is a very simple system in essence, and yet it continues 
to offer novel insights. 

Going by the form of the Li\'enard system derived here, 
it is easy to see that the number of equilibrium points 
will depend on the order of nonlinearity that we may wish
to retain in the equation of the perturbation. In practice,
however, the analytical task becomes formidable with the 
inclusion of every higher order of nonlinearity. Going up 
to the second order, an instability in real time appears 
undeniable, but then we must realize that this conclusion has
been made regarding a purely inviscid and conservative flow. 
Real fluids have viscosity as another important 
physical factor to influence their dynamics. In fact, fluid 
flows are usually affected by both nonlinearity and viscosity,
occasionally as competing effects, and apropos of this point,
we note that for a linearized 
radial perturbation in spherically 
symmetric inflows, viscosity helps in decaying the amplitude 
of the standing waves on globally subsonic solutions~\citep{ray03}. 
So the instability that arises because of nonlinearity can 
very well be tempered by viscosity in the flow~\citep{sb78}. 
Closely related to viscous dissipation, the stability of 
spherically symmetric accretion is expected to be 
affected by turbulence as well~\citep{rb05}. 

A suitable mechanism that favours stability may 
be found in accretion onto black holes, where the coupling 
of the flow with the geometry of space-time acts in the manner 
of a dissipating effect. General relativistic effects have been 
known to enhance the stability of the flow~\citep{ncbr07}.  
And the stability of fluids may also be studied by constraining 
a perturbation to behave like a travelling 
wave~\citep{pso80,rbpla07,ncbr07,sbbr13}. At times, we 
encounter the surprising situation of a fluid flow being 
stable under one type of perturbation, but unstable under 
the effect of another~\citep{rbpla07}. Formally involving 
nonlinearity, all these features merit a close examination.  

\begin{acknowledgments} 
This research has made use of NASA's Astrophysics Data System. 
It was carried out as a part of the {\it National Initiative on 
Undergraduate Science} (NIUS), conducted by Homi Bhabha Centre 
for Science Education, Tata Institute of Fundamental Research, 
Mumbai, India. The authors express their indebtedness to 
J. K. Bhattacharjee, T. K. Das and S. Roy Chowdhury for 
useful comments, and to S. Bobade, A. R. Dhakulkar and 
A. Mazumdar for their support in various academic matters. 
\end{acknowledgments} 

\bibliography{sr2013}
\end{document}